\documentclass[prb,twocolumn]{revtex4}
\usepackage{pstricks}

\begin{document}

\title{An Evolutionary Formalism for Weak Quantum Measurements}
\author{Apoorva Patel}
\email{adpatel@cts.iisc.ernet.in}
\author{Parveen Kumar}
\email{parveenkumar@cts.iisc.ernet.in}
\affiliation{Centre for High Energy Physics,
    Indian Institute of Science, Bangalore 560012}

\maketitle

\hrule\vspace{2mm}
{\noindent\bf
Unitary evolution and projective measurement are fundamental axioms
of quantum mechanics. Even though projective measurement yields one
of the eigenstates of the measured operator as the outcome, there is no
theory that predicts which eigenstate will be observed in which experimental
run. There exists only an ensemble description, which predicts probabilities
of various outcomes over many experimental runs. We propose a dynamical
evolution equation for the projective collapse of the quantum state in
individual experimental runs, which is consistent with the well-established
framework of quantum mechanics. In case of gradual weak measurements, its
predictions for ensemble evolution are different from those of the Born
rule. It is an open question whether or not suitably designed experiments
can observe this alternate evolution.}
\vspace{2mm}\hrule\vspace{2mm}


\section{The Problem}

This talk is about filling a gap in the existing framework of quantum
mechanics. At its heart, quantum mechanics contains two distinct
dynamical rules for evolving a state. One is unitary evolution,
specified by the Schr\"odinger equation:
\begin{equation}
i{d \over dt}|\psi\rangle = H|\psi\rangle ~,~~
i{d \over dt}\rho = [H,\rho] ~.
\end{equation}
It is continuous, reversible and deterministic. The other is the von
Neumann projective measurement, which gives one of the eigenvalues of
the measured observable as the measurement outcome and collapses the
state to the corresponding eigenvector. With $P_i$ denoting the
projection operator for the eigenvalue $\lambda_i$, 
\begin{equation}
|\psi\rangle \longrightarrow P_i |\psi\rangle / |P_i |\psi\rangle| ,~
P_i^2 = P_i = P_i^\dagger ,~ \sum_i P_i = I .
\end{equation}
This change is discontinuous, irreversible and probabilistic in the
choice of ``$i$". It is consistent on repetition, i.e. a second
measurement of the same observable on the same system gives the same
result as the first one.

Both these evolutions, not withstanding their dissimilar properties,
take pure states to pure states. They have been experimentally verified
so well that they are accepted as axioms in the standard formulation of
quantum mechanics. Nonetheless, the formulation misses something: {\it
While the set of projection operators $\{P_i\}$ is fixed by the measured
observable, only one ``$i$" occurs in a particular experimental run, and
there is no prediction for which ``$i$" that would be.}
 
What appears instead in the formulation is the probabilistic Born rule,
requiring an ensemble interpretation for verification. Measurement of an
observable on a collection of identically prepared quantum states gives:
\begin{equation}
\label{Bornrule}
prob(i) = \langle\psi|P_i|\psi\rangle = Tr(P_i\rho) ~,~~
\rho \longrightarrow \sum_i P_i \rho P_i ~.
\end{equation}
This rule evolves pure states to mixed states. All predicted quantities
are expectation values obtained as averages over many experimental runs.
The appearance of a mixed state also necessitates a density matrix
description, instead of a ray in the Hilbert space description for a
pure state.

Over the years, many attempts have been made to combine these two distinct
quantum evolution rules in a single framework. Although the problem of
which ``$i$" will occur in which experimental run has remained unsolved,
progress has been achieved in understanding the ``ensemble evolution" of
a quantum system.

\subsection{Environmental Decoherence}

The system, the measuring apparatus as well as the environment---all are
ultimately made from the same set of fundamental building blocks. With
quantum theory successfully describing the dynamical evolution of all
the fundamental blocks, it is logical to consider the proposition that
the whole universe is governed by the same set of basic quantum rules.
The essential difference between the system and the environment is that
the degrees of freedom of the system are observed while those of the
environment are not. (In a coarse-grained view, unobserved degrees of
freedom of the system can be treated in the same manner as those of the
environment.) All the unobserved degrees of freedom then need to be
``summed over" to determine how the remaining observed degrees of
freedom evolve.

No physical system is perfectly isolated from its surroundings.
Interactions between the two, with a unitary evolution for the whole
universe, entangles the observed system degrees of freedom with the
unobserved environmental degrees of freedom. When the unobserved
degrees of freedom are summed over, a pure but entangled state for
the universe reduces to a mixed state for the system:
\begin{equation}
|\psi\rangle_{SE} \longrightarrow U_{SE} |\psi\rangle_{SE} ~,~~
\rho_S = Tr_E(\rho_{SE}) ~,~~ \rho_S^2 \ne \rho_S ~.
\end{equation}
In general, the evolution of a reduced density matrix is linear, Hermiticity
preserving, trace preserving and positive, but not unitary. Using a complete
basis for the environment $\{|\mu\rangle_E\}$, such a superoperator evolution
can be expressed in the Kraus decomposition form:
\begin{eqnarray}
\label{krausdecomp}
&& \rho_S \longrightarrow \sum_\mu M_\mu \rho_S M_\mu^\dagger ~,\\
&& M_\mu = {}_E\langle\mu | U_{SE} | 0\rangle_E ~,~~
   \sum_\mu M_\mu^\dagger M_\mu = I ~.\nonumber
\end{eqnarray}
This description explains the probabilistic ensemble evolution of quantum
mechanics in a language similar to that of classical statistical mechanics.
But it still has no mechanism to explain the projective collapse of a
quantum state. (Ensemble averaging is often exchanged for ergodic time
averaging in equilibrium statistical mechanics, but that option is not
available in unitary quantum mechanics.)

Generically the environment has a much larger number of degrees of freedom
than the system. Then, in the Markovian approximation which assumes that
information leaked from the system does not return, the evolution of
the reduced density matrix can be converted to a differential equation.
With the expansion
\begin{equation}
M_0 = I + dt (-iH + K) + \ldots ~,~~
M_{\mu\ne0} = \sqrt{dt}L_\mu + \ldots ~,
\end{equation}
Eq.(\ref{krausdecomp}) leads to the Lindblad master equation
\cite{Lindblad,GKS}:
\begin{eqnarray}
\label{mastereqn}
{d\over dt}\rho &=& i[\rho,H] + \sum_{\mu}{\cal L}[L_\mu]\rho ~,\\
{\cal L}[L_\mu]\rho &=& L_\mu\rho L_\mu^\dagger
  - {1\over2} \rho L_\mu^\dagger L_\mu - {1\over2} L_\mu^\dagger L_\mu\rho ~.
\nonumber
\end{eqnarray}
The terms on the r.h.s. involving sum over $\mu$ modify the unitary
Schr\"odinger evolution, while $Tr(d\rho/dt)=0$ preserves the total
probability. When $H=0$, the fixed point of the evolution is a diagonal
$\rho$, in the basis that diagonalises $\{L_\mu\}$. This preferred basis
is determined by the system-environment interaction. (When there is no
diagonal basis for $\{L_\mu\}$, the evolution leads to equipartition,
i.e. $\rho \propto I$.) Furthermore, the off-diagonal components of
$\rho$ decay, due to destructive interference among environmental
contributions with varying phases, which is known as decoherence.

This modification of a quantum system's evolution, due to its coupling
to unobserved environmental degrees of freedom, provides the correct
ensemble interpretation, and a quantitative understanding of how the
off-diagonal components of $\rho$ decay \cite{Zeh,Wiseman}. Still the
quantum theory is incomplete, and we need to look further to solve the
``measurement problem" till it can predict the outcome of a particular
experimental run.

\subsection{Going Beyond}

A wide variety of theoretical approaches have been proposed to get
around the quantum measurement problem. Some of them are physical,
e.g. introduction of hidden variables with novel dynamics, and
breakdown of quantum rules due to gravitational interactions.
Some others are philosophical, e.g. questioning what is real and what
is observable, in principle as well as by human beings with limited
capacity. Given the tremendous success of quantum theory, realised
with a ``shut up and calculate" attitude, and the stringent constraints
that follow, none of the theoretical approaches have progressed to the
level where they can be connected to readily verifiable experimental
consequences.

Perhaps the least intrusive of these approaches is the ``many worlds
interpretation" \cite{Everett}. It amounts to assigning a distinct
world (i.e. an evolutionary branch) to each probabilistic outcome,
while we only observe the outcome corresponding to the world we live in.
(It is amusing to note that this discussion meeting is being held in a
place where the slogan of the department of tourism is ``One state, Many
worlds" \cite{Karnataka}.) Such an entanglement between the measurement
outcomes and the observers does not violate any quantum principle,
although the uncountable proliferation of evolutionary branches it
supposes is highly ungainly. Truly speaking, the many worlds
interpretation bypasses the measurement problem instead of solving it.

With the technological progress in making quantum devices, we need a
solution to the measurement problem, not only for formal theoretical
reasons, but also for increasing accuracy of quantum control and feedback
\cite{Wiseman}. A practical situation is that of the weak measurement
\cite{Aharonov}, typically realised using a weak system-apparatus coupling,
where information about the measured observable is extracted from the
system at a slow rate. Such a stretching out of the time scale allows
one to monitor how the system state collapses to an eigenstate of the
measured observable, and to track properties of the intermediate states
created along the way by an incomplete measurement. Knowledge of what
really happens in a particular experimental run (and not the ensemble
average) would be invaluable in making quantum devices more efficient
and stable.

\section{A Way Out}

Let us assume that the projective measurement results from a continuous
geodesic evolution of the initial quantum state to an eigenstate $|i\rangle$
of the measured observable:
\begin{equation}
|\psi\rangle \longrightarrow Q_i(s)|\psi\rangle / |Q_i(s)|\psi\rangle| ,~
Q_i(s) = (1-s)I + sP_i ,
\end{equation}
where the dimensionless parameter $s\in[0,1]$ represents the
``measurement time". The density matrix then evolves as,
maintaining $Tr(\rho)=1$,
\begin{equation}
\rho \longrightarrow
     { (1-s)^2\rho + s(1-s)(\rho P_i + P_i\rho) + s^2 P_i\rho P_i 
     \over  (1-s)^2 + (2s-s^2)Tr(P_i\rho) } ~.
\end{equation}
Expansion around $s=0$ yields the differential equation:
\begin{equation}
\label{projmeas}
{d\over ds}\rho = \rho P_i + P_i\rho - 2\rho~Tr(P_i\rho)
                = \{\rho,P_i\} - \rho~Tr(\{\rho,P_i\}) ~.
\end{equation}
This simple equation describing an individual quantum trajectory has
several remarkable properties. We can explore them, putting aside the
argument that led to the equation. Explicitly,

\begin{itemize}
\item
In addition to maintaining $Tr(\rho)=1$, the nonlinear evolution preserves
pure states. $\rho^2=\rho$ implies $\rho P_i \rho = \rho~Tr(P_i\rho)$,
and then
\begin{equation}
\label{evolpure}
{d\over ds}\big(\rho^2 - \rho\big)
= \rho{d\over ds}\rho + \big({d\over ds}\rho\big)\rho - {d\over ds}\rho = 0 ~.
\end{equation}
For pure states, with $\langle\psi|{d\over ds}|\psi\rangle = 0$,
we can also write
${d\over ds}|\psi\rangle = (P_i - \langle\psi|P_i|\psi\rangle) |\psi\rangle$.
So the component of the state along $P_i$ grows at the expense of the other
orthogonal components.

\item
Each projective measurement outcome is the fixed point of the
deterministic evolution:
\begin{equation}
{d\over ds}\rho = 0 \quad{\rm at}\quad
\rho_i^* = P_i\rho P_i/Tr(P_i\rho) ~.
\end{equation}
The fixed point nature of the evolution makes the measurement consistent
on repetition. Note that one-dimensional projections satisfy
$P_i\rho P_i = P_i~Tr(P_i\rho)$.

\item
In a bipartite setting, the complete set of projection operators can be
labeled as $\{P_i\} = \{P_{i_1} \otimes P_{i_2}\}$, with $\sum_i P_i = I$.
Since the evolution is linear in the projection operators, a partial trace
over the unobserved degrees of freedom produces the same equation (and
hence the same fixed point) for the reduced density matrix for the system.
Purification is thus a consequence of the evolution; for example, a qubit
state in the interior of the Bloch sphere evolves to the fixed point on
its surface.

\item
At the start of measurement, we expect the parameter $s$ to be
proportional to the system-apparatus interaction, $s\sim||H_{SA}||t$.
To understand the approach towards the fixed point, let
$\tilde{\rho}\equiv\rho-P_i$ for a one-dimensional projection,
which satisfies
\begin{equation}
{d\over ds}\tilde{\rho} = \tilde{\rho}P_i + P_i\tilde{\rho}
      - 2\tilde{\rho} - 2\rho~Tr\big(P_i\tilde{\rho}\big) ~.
\end{equation}
It follows that towards the end of measurement $s\rightarrow\infty$,
and convergence to the fixed point is exponential, with
$||\tilde{\rho}|| \sim e^{-2s}$, similar to the charging of a capacitor.
\end{itemize}

\noindent
These properties make Eq.(\ref{projmeas}) a legitimate candidate for
describing the collapse of a quantum state during projective measurement.
It represents a superoperator that preserves Hermiticity, trace and
positivity, but is nonlinear. Because of its non-stochastic nature,
it can model the single quantum trajectory specific to a particular
experimental run.

Although Eq.(\ref{projmeas}) does fill a gap in solving the measurement
problem, with the preferred basis $\{P_i\}$ fixed by the system-apparatus
interaction, we still need a separate criterion to determine which $P_i$
will occur in a particular experimental run. This is a situation 
reminiscent of spontaneous symmetry breaking \cite{Ghose}, where a tiny
external field (with a smooth limit to zero) picks the direction, and
the evolution is unique given that direction (stability of the direction
depends on the thermodynamic size of the system). We do not have a
prescription for such a choice, also referred to as a ``quantum jump".
The stochastic ensemble interpretation of quantum measurements is reproduced,
as per the Born rule, when a particular $P_i$ is picked with probability
$Tr(P_i\rho(s=0))$.

\subsection{Combining Trajectories}

The probabilistic Born rule for measurement outcomes, Eq.(\ref{Bornrule}),
is rather peculiar despite being tremendously successful. The reason is
that the probabilities are determined by the initial state $\rho(s=0)$,
and do not depend on the subsequent evolution of the state $\rho(s\ne0)$.
Any attempt to describe projective measurement as continuous evolution
would run into the problem that the system would have to remember its
state at the instant the measurement started until the measurement is
complete. This is a severe constraint for any theory of weak measurement,
where the measurement time scale is stretched out, and we can rightfully
question whether the Born rule would hold in such a case.

It is possible to reconcile the Born rule with continuous projective
measurements, by invoking retardation effects arising from special
relativity for the speed of information travel between the system and
the apparatus. Then the Born rule will be followed by sudden impulsive
measurements with a duration shorter than the retardation time, but it
may be violated by gradual weak measurements with a duration longer
than the retardation time. We look beyond the Born rule satisfying
possibility that the evolution trajectory corresponding to $P_i$ is
chosen at the start of the measurement and remains unaltered thereafter,
in order to look for more general evolutionary choices that may be
suitable for weak measurements.

Let $w_i$ be the probability weight of the evolution trajectory for $P_i$,
with $w_i\ge0$ and $\sum_i w_i = 1$. We have $w_i(s=0)=\rho_{ii}(s=0)$ in
accordance with the Born rule, while $w_i(s\ne0)$ are some functions of
$\rho(s)$. Then the trajectory averaged evolution of the density matrix
during measurement is given by:
\begin{equation}
\label{trajavmeas}
{d\over ds}\rho = \sum_i w_i [\rho P_i + P_i\rho - 2\rho~Tr(P_i\rho)] ~.
\end{equation}
It still preserves pure states, as per Eq.(\ref{evolpure}).
In terms of a complete set of projection operators, we can decompose
$\rho = \sum_{jk} P_j \rho P_k$. The projected components evolve as
\begin{equation}
{d\over ds}\big(P_j \rho P_k\big)
= P_j \rho P_k \big( w_j + w_k - 2\sum_i w_i~Tr(P_i\rho) \big) ~.
\end{equation}
This evolution obeys the identity, independent of the choice of $\{w_i\}$,
\begin{eqnarray}
{2\over P_j \rho P_k}{d\over ds}\big(P_j \rho P_k\big)
&=& {1\over P_j \rho P_j}{d\over ds}\big(P_j \rho P_j\big) \nonumber\\
&+& {1\over P_k \rho P_k}{d\over ds}\big(P_k \rho P_k\big) ~,
\end{eqnarray}
with the consequence that the diagonal projections of $\rho$ completely
determine the evolution of all the off-diagonal projections. The diagonal
projections are all non-negative, $P_j \rho(s) P_j = d_j(s) P_j$ with
$d_j\geq0$, and we obtain:
\begin{equation}
\label{offdiagevol}
P_j \rho(s) P_k = P_j \rho(0) P_k
\left[ {d_j(s)~d_k(s) \over d_j(0)~d_k(0)} \right]^{1/2} ~.
\end{equation}
In particular, phases of the off-diagonal projections $P_j \rho P_k$ do not
evolve, in sharp contrast to what happens during decoherence. Also, their
asymptotic values, i.e. $P_j \rho(s\rightarrow\infty) P_k$, may not vanish,
whenever more than one diagonal $P_j \rho(s\rightarrow\infty) P_j$ remain
nonzero. In a sense, decohering measurements select the Cartesian components
of the quantum state in the eigenbasis provided by the system-apparatus
interaction and lose information about the angular coordinates, while the
collapse equation selects the radial components of the quantum state around
the measurement fixed points leaving the angular components unchanged.
Mathematically speaking, both measurement schemes are consistent.

It is easily seen that when all the $w_i$ are equal, no information is
extracted from the system by the measurement and $\rho$ does not evolve.
More generally, the diagonal projections evolve according to:
\begin{equation}
\label{diagevol}
{d\over ds}d_j = 2d_j \big( w_j - \sum_i w_i d_i \big)
               = 2d_j \big(\sum_{i\ne j} (w_j - w_i)d_i \big) .
\end{equation}
Here, with $\sum_i d_i=1$, $\sum_i w_i~d_i \equiv \overline{w}$ is the
weighted average of $\{w_i\}$. Clearly, the diagonal projections with
$w_j > \overline{w}$ grow and the ones with $w_j < \overline{w}$ decay.
Any $d_j$ that is zero initially does not change, and the evolution is
therefore restricted to the subspace spanned by all the $P_j\rho(s=0)P_j
\ne 0$. Also, all the measured observable eigenstates, i.e. $\rho=P_j$
with $d_j=1$, are fixed points of the evolution. These features are
stable under small perturbations of the density matrix.

Other fixed points of Eq.(\ref{diagevol}) correspond to ``degenerate"
situations where some of the $w_j$ (say $n>1$ in number) are equal and
all the others vanish, i.e. $w_j \in \{0,{1\over n}\}$. These fixed points
are unstable under asymmetric perturbations that lift the degeneracy.
It may be that other terms in the evolution Hamiltonian, which have been
ignored throughout in our measurement description and whose contribution
would have to be added to Eq.(\ref{trajavmeas}) in describing complete
evolution of the system, can stabilise them, and make the evolution
converge towards an $n$-dim degenerate subspace.

An appealing choice for the trajectory weights is the ``instantaneous Born
rule", i.e. $w_j = w_j^B \equiv Tr(P_j\rho(s))$ throughout the measurement
process. That avoids logical inconsistency in weak measurement scenarios,
where one starts the measurement, pauses somewhere along the way, and then
restarts the measurement. In this situation, the trajectory averaged
evolution is:
\begin{equation}
{d\over ds}\big(P_j \rho P_k\big)
= P_j \rho P_k (w_j^B + w_k^B - 2\sum_i (w_i^B)^2) ~.
\end{equation}
This evolution converges towards the $n-$dimensional subspace specified
by the dominant diagonal projections of the initial $\rho(s=0)$. It is
deterministic and does not follow Eq.(\ref{Bornrule}). The measurement
result remains consistent under repetition though.

The evolution can be made stochastic, in a manner similar to the Langevin
equation, by adding noise to the weights $w_i$ while still retaining
$\sum_i w_i=1$. The weak measurement process is expected to contribute
such a noise \cite{Korotkov,Vijay,Murch}. The resultant evolution, and
its dependence on the magnitude of the noise, needs to be investigated.

\subsection{Relation to the Master Equation}

The master equation is obtained assuming that the environmental degrees
of freedom are not observed, and hence are summed over. On the other hand,
the degrees of freedom corresponding to the measured observable are
observed in any measurement process with a definite outcome, and cannot be
summed over. In analysing the measurement process, we need to keep track
of only those degrees of freedom of the apparatus that have a one to one
correspondence with the system's eigenstates, and the rest can be kept
aside. The crucial difference between the states of the system and the
apparatus is that the system can be in a superposition of the eigenstates
but the apparatus has to end up in one of the pointer states only (and
not their superposition).

In the traditional description, at the start of the measurement,
the joint state of the system and the apparatus can be chosen to be
$\sum_i c_i |i\rangle_S |0\rangle_A$, with $\sum_i |c_i|^2 = 1$.
The system-apparatus interaction then unitarily evolves it to the
entangled state $ \sum_i c_i |i\rangle_S |\tilde{i}\rangle_A$.
This evolution is a controlled unitary transformation (and not a copy
operation). The preferred measurement basis is the Schmidt decomposition
basis, ensuring a perfect correlation between the system eigenstate
$|i\rangle_S$ and the measurement pointer state $|\tilde{i}\rangle_A$.
In particular, the reduced density matrices of the system and the apparatus
are identical at this stage. Thereafter, the state collapse picks one of
the components $|i\rangle|\tilde{i}\rangle$, without losing the perfect
correlation.

The algebraic structure of the collapse equation, Eq.(\ref{projmeas}), 
is closely related to that of the master equation, Eq.(\ref{mastereqn}).
Expansion of Eq.(\ref{projmeas}) around the fixed point $\rho=P_i$ gives,
with ${\cal L}[P_i]P_i=0$,
\begin{equation}
{d\over ds}\tilde{\rho} = 2{\cal L}[P_i]\tilde{\rho}
           - 2 P_i\tilde{\rho}P_i - 2(1-P_i)\tilde{\rho}(1-P_i)
           - 2\tilde{\rho}~Tr(\tilde{\rho}P_i) ~.
\end{equation}
The term ${\cal L}[P_i]\tilde{\rho} = {\cal L}[P_i]\rho$ on the r.h.s.
decouples $P_i\rho P_i$ from the rest of the density matrix by making the
off-diagonal components ($P_i\rho P_j$ and $P_j\rho P_i$) decay, but does
not alter the diagonal components. The next two terms on the r.h.s. make
the diagonal components of $\tilde{\rho}$ decay, leading the evolution
to the fixed point. The last term on the r.h.s. is of higher order in
$\tilde{\rho}$.

The Lindblad operators also satisfy the relation,
\begin{equation}
{\cal L}[\rho]P_i + {\cal L}[P_i]\rho = {\cal L}[\tilde{\rho}]P_i
                                      = O(\tilde{\rho}^2) ~.
\end{equation}
${\cal L}[P_i]\rho$ is the influence of the apparatus pointer state on
the system density matrix, while ${\cal L}[\rho]P_i$ is the influence
of the system density matrix on the apparatus pointer state. For pure
states, the collapse equation is just
\begin{equation}
{d\over ds}\rho = -2{\cal L}[\rho] P_i ~.
\end{equation}
These expressions suggest an inverse relationship between the processes
of decoherence and collapse. Such an action-reaction relationship can
follow from a conservation law. Initially, the combined system-apparatus
state evolves unitarily, establishing perfect correlation and without
any decoherence. The subsequent new description would be that during
the measurement process, when ${\cal L}[\rho]P_i$ decoheres the apparatus
pointer state $P_i$ (it cannot remain in superposition), there is an equal
and opposite effect $-{\cal L}[\rho]P_i$ on the system density matrix $\rho$,
resulting in the state collapse.

\subsection{The Qubit Case and Some Tests}

All the previous results can be expressed in a considerably simpler form
in case of the smallest quantum system, i.e. the 2-dimensional qubit with
$|0\rangle$ and $|1\rangle$ as the measurement basis vectors. Evolution
of the density matrix during the measurement, Eqs.(\ref{diagevol}) and
(\ref{offdiagevol}), is given by:
\begin{equation}
\label{oneqevold}
{d\over ds}\rho_{00}(s) = 2(w_0-w_1)~\rho_{00}(s)~\rho_{11}(s) ~,
\end{equation}
\begin{equation}
\label{oneqevolod}
\rho_{01}(s) = \rho_{01}(0) \left[ {\rho_{00}(s)~\rho_{11}(s)
               \over \rho_{00}(0)~\rho_{11}(0)} \right]^{1/2} ~.
\end{equation}
Selecting the trajectory weights as addition of Gaussian white noise
to the instantaneous Born rule results in:
\begin{equation}
\label{oneqweight}
w_0 - w_1 = \rho_{00} - \rho_{11} + \xi ~.
\end{equation}
The same evolution has been obtained by Korotkov \cite{Korotkov},
using the Bayesian measurement formalism for a qubit. Our analysis is
more general and is applicable to any quantum system.

For a single transmon qubit undergoing Rabi oscillations, perturbations
caused by its weak measurement have been observed, and then successfully
cancelled by a measurement result dependent feedback shift of the Rabi
frequency \cite{Vijay}, all consistent with the description provided by
Eqs.(\ref{oneqevold}-\ref{oneqweight}). We have verified the same
behaviour for two qubit systems using numerical simulations based on
Eqs.(\ref{diagevol}) and (\ref{offdiagevol}), and trajectory weights
chosen analogous to Eq.(\ref{oneqweight}), e.g. perturbations due to weak
measurements of $J_z \otimes J_z$ on a Bell state undergoing joint Rabi
oscillations can be cancelled by a measurement result dependent Rabi
frequency shift. We could not make this cancellation strategy work, however,
with simultaneous measurement of more than one commuting operators and
independent shifts of individual Rabi frequencies.

\section{Open Questions}

Our proposed collapse equation, Eq.(\ref{projmeas}), is quadratically
nonlinear. Nonlinear Schr\"odinger evolution is inappropriate in quantum
mechanics, because it violates the well-established superposition principle.
Nonlinear superoperator evolution for the density matrix is also avoided in
quantum mechanics, because it conflicts with the probability interpretation
for mixtures of density matrices. Nevertheless, nonlinear quantum evolutions
need not be unphysical, and have been invoked in attempts to solve the
measurement problem \cite{GMH,Penrose}. Eq.(\ref{projmeas}) can be a
valuable intermediate step in such attempts to interpret collapse as a
consequence of some unknown underlying dynamics. It is definitely worth
keeping in mind that non-abelian gauge theories and general relativity
are examples of well-established theories with quadratic nonlinearities
in their dynamical equations.

Irrespective of the underlying dynamics that may lead to the collapse
equation, Eq.(\ref{projmeas}), it is worthwhile to test it at its face
value. It readily produces an eigenstate of the measured observable as
the measurement outcome, but predictions of its ensemble version,
Eq.(\ref{trajavmeas}), are not stochastic and do not reproduce the Born
rule. So to judge its validity, it is imperative to ask the question:
{\it Do quantum systems exhibit this alternate evolution, and if so
under what conditions?}
Experimental tests would require determination of the density matrix
evolution, in presence of weak measurements and with highly suppressed
decoherence effects. Such tests are now technologically feasible!
It is indeed possible to generalise and extend the Bayesian measurement
formalism tests for a single qubit, by R. Vijay et al. \cite{Vijay} and
by K.W. Murch et al. \cite{Murch}, to larger quantum systems. They would
clarify what trajectory weights $w_i$ (including stochastic noise) are
appropriate for describing ensemble dynamics of weak measurements.

Finally, it is useful to note that, if the fixed point collapse dynamics
of Eq.(\ref{projmeas}) can be realised in practice, it would provide an
unusual strategy for quantum error correction. After encoding the logical
Hilbert space as a suitable subspace of the physical Hilbert space, one
only has to perform measurements in an eigenbasis that separates the logical
subspace and the error subspace as orthogonal projections. The state would
then return towards the logical subspace as long as its projection on the
logical subspace is larger than that on the error subspace, even if no
feedback operations based on the measurement results are carried out!

\noindent{ACKNOWLEDGEMENTS.}
We thank Michel Devoret and Rajamani Vijayaraghavan for helpful
conversations. The slogan of the Karnataka Tourism Department was
pointed out to us by Raymond Laflamme. PK is supported by a CSIR
research fellowship.

\end{document}